\documentclass[10pt,twocolumn,letterpaper]{article}
\usepackage{enumitem}
\usepackage{array}
\usepackage{cvpr}
\usepackage{multirow}
\usepackage{times}
\usepackage{epsfig}
\usepackage{graphicx}
\usepackage{amsmath}
\usepackage{amssymb}
\usepackage{microtype} 

\newcolumntype{C}[1]{>{\centering\let\newline\\\arraybackslash\hspace{0pt}}m{#1}}
\newcolumntype{R}[1]{>{\raggedleft\let\newline\\\arraybackslash\hspace{0pt}}m{#1}}

\usepackage[breaklinks=true,bookmarks=false]{hyperref}
\usepackage{array}

\newcommand{\superscript}[1]{\ensuremath{^{\textrm{#1}}}}
\def\sharedaffiliation{\end{tabular}\newline\begin{tabular}{c}}
\def\upf{\superscript{2}}
\def\yhoo{\superscript{1}}
\def\unito{\superscript{3}}

\makeatletter
\newcommand{\thickhline}{%
    \noalign {\ifnum 0=`}\fi \hrule height 1pt
    \futurelet \reserved@a \@xhline
}
\makeatother
 \cvprfinalcopy 

\def\cvprPaperID{1964} 

\ifcvprfinal\fi

\begin{document}

\title{6 Seconds of Sound and Vision: Creativity in Micro-Videos}


\author{
\hspace{-0.5cm} Miriam Redi\yhoo\\
\and
\hspace{-0.5cm} Neil O'Hare\yhoo\\
\and
\hspace{-0.5cm} Rossano Schifanella\unito\superscript{,}\thanks{This work has been performed when the author was a Visiting Scientist at Yahoo Labs, Barcelona, within the framework of the FREP grant.} \\
\and
\hspace{-0.5cm} Michele Trevisiol\upf\superscript{,}\yhoo\\
\and
\hspace{-0.5cm} Alejandro Jaimes\yhoo\\
%
\sharedaffiliation
{\yhoo}Yahoo Labs, Barcelona, Spain {\tt \small \{redi,nohare,ajaimes\}@yahoo-inc.com}  \\
{\upf}Universitat Pompeu Fabra, Barcelona, Spain {\tt \small \{trevisiol\}@acm.org} \\
{\unito}Universit\`a degli Studi di Torino, Torino, Italy {\tt \small \{schifane\}@di.unito.it}
}

\maketitle

\begin{abstract}
The notion of creativity, as opposed to related concepts such as beauty or interestingness, has not been studied from the perspective of automatic analysis of multimedia content. Meanwhile, short online videos shared on social media platforms, or micro-videos, have arisen as a new medium for creative expression. In this paper we study creative micro-videos in an effort to understand the features that make a video creative, and to address the problem of automatic detection of creative content. Defining creative videos as those that are novel and have aesthetic value, we conduct a crowdsourcing experiment to create a dataset of over 3,800 micro-videos labelled as creative and non-creative. We propose a set of computational features that we map to the components of our definition of creativity, and conduct an analysis to determine which of these features correlate most with creative video. Finally, we evaluate a supervised approach to automatically detect creative video, with promising results, showing that it is necessary to model both aesthetic value and novelty to achieve optimal classification accuracy.
\end{abstract}

\section{Introduction} \label{sec:introduction}



Short online videos, or \textit{micro-videos}, have recently emerged as a new form of user-generated content on social media platforms such as Vine, Instagram, and Facebook\footnote{\tiny \url{http://vine.co}, \url{http://instagram.com}, \url{http://facebook.com}}. 
The  Vine platform, in particular, has become associated with the notion of creativity, as it was launched with the goal of allowing users to create 6-second videos  whose time constraint ``inspires creativity''\footnote{\tiny {\url{http://blog.vine.co/post/55514427556/introducing-vine}}}.
Some commentators have even claimed of Vine in particular that ``its constraints were allowing digital videos to take on entirely new forms''\footnote{\tiny {\tiny \url{https://medium.com/art-technology/a4433fb334f}}}, and interest in Vine videos has prompted the creation of a specific 6-second film category at major film festivals such as the Tribeca Film Festival in New York. 



Not all micro-videos uploaded on social media platforms are creative in nature (1.9\% of randomly sampled videos were annotated as creative in our study), and 
quality can vary widely. This motivates the need for automatic approaches to detect and rank the best, and in particular the most \textit{creative}, micro-video content on social media platforms. 
Such applications can increase the visibility of video authors, and replace or augment current features of social-media platforms such as ``Editors Picks'', which showcases the best content on Vine.

\textit{Micro-videos} provide a unique opportunity to address the study of audio-visual creativity using computer vision and audio analysis techniques. 
The very short nature of these videos 
means that we can analyze them at a micro-level. Unlike short video sequences within longer videos, 
the 
information required to understand a micro-video 
is contained within the video itself. 
This allows us to study audio-visual creativity at a fine-grained level, helping us to understand what, exactly, constitutes creativity in micro-videos. 

In this paper we study the audio-visual features of \textit{creative} vs \textit{non-creative} videos\footnote{Throughout the paper we will use the word "video" to refer to "micro-videos" of a few seconds} and present a computational framework to automatically classify these categories. In particular, we conduct a crowdsourcing experiment to annotate over 3,800 Vine videos, using as guidelines: (1) a widely accepted definition of creative artifacts as those that are \textit{novel} and \textit{valuable}, and (2) insights from the philosophy of aesthetics about the judgements of aesthetic value (i.e. sensory, emotional/affective, and intellectual). We go on to use this dataset to study creative \textit{micro-videos} and to evaluate approaches to automatic detection of creative micro-videos.


The main contributions of this paper are:
\begin{itemize}
\item We create a new dataset of creative \textit{micro-videos}, and make the vine video ids and annotations publicly available to the research community\footnote{available for download at: \url{ http://di.unito.it/vinecvpr14}}.
\item We propose and implement a new set of  features to model the novelty and aesthetic value of \textit{micro-videos}.
\item We analyze the extent to which each of these features, and other existing features, correlate with creativity, giving insights into the audio-visual features most associated with creative video. We also classify videos as creative/non-creative, with promising results, and we show that combining aesthetic value and novelty features gives highest accuracy.
\end{itemize}
Unlike previous work in computational aesthetics \cite{datta,interesting}, which mainly focuses on assessing visual beauty using compositional features, we explore here the more complex and subtle concept of \textit{creativity}. Focusing on creative content allows us to analyze audio-visual content from a  different perspective, allowing us to model the fact that creative content is not always the most beautiful-looking (in the conventional sense) or visually interesting. To the best of our knowledge, this is the first work to address creativity in micro-videos.
In the next Section we present related work, and we define video creativity in Section \ref{sec:creativity}. 
In Section \ref{sec:dataset} we describe a crowdsourced annotation of Vine videos. Section \ref{sec:features} presents computational features for modeling creativity. In Section \ref{sec:results} we correlate these features with, and evaluate the automatic classification of, creative content. We conclude in Section \ref{sec:conclusions}.
\section{Related Work} \label{sec:relatedwork}
Our work is closely related to  computational approaches to studying concepts such as beauty \cite{datta}, interestingness \cite{interesting}, memorability \cite{isola2011makes}, or emotions \cite{emotions}. 
In particular, we are influenced by recent work in computational aesthetics for the automatic  assessment of visual beauty. The earliest work \cite{datta,ke2006design} distinguishes between high-quality (professional) and low-quality (amateur) photos based on features inspired by photographic rules, with applications in image quality enhanchement \cite{bhattacharya2010framework} and automatic aesthetic feedback for photographers \cite{yao2012oscar}. Nishiyama \etal \cite{nishiyama2011aesthetic} propose more complex visual features based on color harmony, and combine them with low-level features for aesthetic image classification. Other work has  investigated generic local features for modeling beauty, showing that they outperform systems based on compositional features \cite{marchesotti2011assessing}. 
Several researchers have included the semantics of the image in the aesthetic evaluation, labeling images according to their scene content 
and building category-based beauty models \cite{luo2011content,murray2012ava}. 

The main difference between visual aesthetic research and our work is that the notion of \textit{creativity} is more complex than visual photographic beauty, in addition to the fact that we also focus on audio. 
We argue that creative videos may not be always considered `beautiful' in the conventional sense, and may even be `ugly'. While we incorporate and re-elaborate many of the mentioned approaches for detecting creative videos, by using sensory (including aesthetic), and visual affect features, we also design a new set of features to model audio-visual creativity.

Moreover, while much related work focuses on still images, in our work we analzye \textit{video} data, and we build specific video features for micro-videos. The few previous works on video aesthetics build video features based on professional movie aesthetics \cite{Chung:EECS-2012-172, bhattacharya2013towards}, or simply aggregate frame-level features \cite{moorthy2010towards}, with limited success. 

Also different from much of the work in computational aesthetics, we use a croudsourced groundtruth, allowing us to create a high quality labelled dataset using a set of annotation guidelines tailored for creativity. Crowdsourcing was previously used to build a corpus for image memorability \cite{isola2011makes}, but most computational aesthetics research exploits online professional photo websites such as \textit{dpchallenge.com} \cite{datta,ke2006design,interesting,murray2012ava,luo2011content}, \textit{photo.net} \cite{luo2011content}, or Flickr \cite{interesting}. 
\section{Defining Video Creativity} \label{sec:creativity}
Although the precise definition of creativity has been the subject of debate in many disciplines, one of the most common observations is that creativity is connected with imagination and innovation, and with the production of novel, unexpected solutions to problems \cite{newell1959processes}. However, \textit{``All who study creativity agree that for something to be creative, it is not enough for it to be novel: it must have value, or be appropriate to the cognitive demands of the situation''} \cite{weisberg1993creativity}, an idea that is shared by many researchers \cite{higgins1999applying, mumford2003have, weisberg1993creativity}. 
Based on these observations, we define a creative artifact as one that is \emph{novel} (surprising, unexpected) and has \emph{value}.

As applied to micro-videos, we interpret by \emph{novelty} that the video is unique in a significant way, or that it expresses ideas in an unexpected or surprising manner.
%
\emph{Value} is a more complex notion, however, and in this context it is best equated with aesthetic value. 
Most definitions of aesthetic value incorporate the maxim that beauty is in the eye of the beholder: Immanuel Kant, for example, in his \textit{Critique of Judgement}\cite{kant1987critique}, argues that aesthetic judgements involve an emotional response (\eg, pleasure) to a sensory input (i.e. the audio-visual signal from the video) that also provokes ``reflective contemplation''. At the risk of oversimplifying, judgements of aesthetic value involve \textit{sensory}, \textit{emotional} and \textit{intellectual} components. 

In the following sections, we will use this definition to: (1) provide a definition of creative video as part of our guidelines for crowd workers to annotate videos as creative or non-creative (Section \ref{sec:dataset}), and (2) inform our choice of computational features for modeling creative videos.



%

\noindent
\section{Dataset} \label{sec:dataset}
%
%
To create a corpus of micro-videos annotated as \textit{creative}, we first identified a set of candidate videos that were likely to be creative. This was necessary because our preliminary analysis showed that only a small fraction of videos are creative, meaning that a random sampling would need an extremely large annotation effort to collect a reasonable number of positive creative videos to analyze. With this in mind, we defined a set of sampling criteria likely to return creative videos. We started by sampling 4,000 videos. Specifically, we took (a) $1,000$ videos annotated with hashtags that were associated to creative content by 3 different blogs about Vine: {\it \#vineart, \#vineartist, \#artwork, and \#vineartgallery} (b) $200$ videos \textsl{mentioned} in $16$ articles about Vine creativity on social media websites, (c) $2,300$ videos authored by the $109$ creators of the videos identified in criteria \textit{b}, based on the assumption that these authors are likely to author other creative micro-videos, and (d) 500 randomly selected videos from the Vine streamline, for the purpose of estimating the true proportion creative videos on Vine. The results of the labeling experiment summarized in Table~\ref{tab:sources_results} confirm the validity of this sampling strategy: while only $1.9\%$ of the random sample has been labeled as creative (D-100), our sampling strategy yielded 25\% creative videos, giving a corpus that is large enough to be useful. In total, after discarding invalid urls, we annotated 3,849 candidate videos, created and shared between November 2012 to July 2013.
%
%
%
%

We annotate these videos using Crowdflower\footnote{\url{http://www.crowdflower.com}}, a large crowdsourcing platform. 
To ensure quality annotations, the platform enables the definition of \emph{Gold Standard} data where workers are assigned a subset of pre-labelled `jobs', allowing the known true label to be compared against the contributor label. This mechanism allows worker performance to be tracked, and can ensure that only judgements coming from competent contributors are considered. It also presents an opportunity to give feedback to workers on how to improve their annotations in response to incorrect answers.

In the experiment, a contributor looks at a 6-second video and judges if it is creative. 
According to Section~\ref{sec:creativity}, a creative video is defined as a video that: (1) has aesthetic value, or evokes an emotion (happy, sad, angry, funny, etc), and (2) has interesting or original/surprising video/audio technique. The worker is advised to listen to the audio, and can watch a pair of exemplar \emph{creative} and \emph{non-creative} videos before performing the job. After watching the target video the contributor answers the question \emph{``Is this video creative?''} with ``positive'', ``negative'' or ``don't know''. In the first two cases, the user can give more details of the motivation of their choice according to the criteria in Table~\ref{table:criteria}, 
phrased in a simple language appropriate to crowdsourcing platforms, where workers typically do not take time to read complex definitions and guidelines~\cite{mason2011}. To ensure that the job could be easily understood by crowd workers, in a preliminary survey we  collected feedback on the interface from 15 volunteers. 

\begin{table}[htdp]

\resizebox{\linewidth}{!}{
\begin{tabular}{|c|l|l|lll|}
\hline

\multirow{4}{*}{\begin{minipage}{0.5in}\textbf{Aesthetic Value}\end{minipage}} & \multirow{2}{*}{\textit{Sensory}} & The audio is appealing/striking \\ 
& & The visuals are appealing/striking\\ \cline{2-3}
& \textit{Emotional} &The video evokes an emotion \\ \cline{2-3}
& \textit{Intellectual} &The video suggests interesting ideas \\ \hline

\multirow{3}{*}{\textbf{Novelty}} & \multicolumn{2}{l|}{The audio is original/surprising}   \\ 
& \multicolumn{2}{l|}{The visuals are original/surprising} \\
& \multicolumn{2}{l|}{The story or content is original/surprising} \\ \hline

\end{tabular}}
\caption{Criteria for labeling a video as creative}
\label{table:criteria}
\end{table}

%
%

The experiment ran for 5 days and involved 285 active workers (65 additional workers were discarded due to the low quality of their annotations) located in USA (88\%),  United Kingdom (8\%), and Germany (4\%).\footnote{Additional demographic information was not available.} No time constraint was set on the task, 
and each video was labeled by 5 independent workers. 
The final annotations reached a level of 84\% worker agreement (82\% for creative, 85\% for non-creative), which we consider high for this subjective task. Looking at per-video agreement, summarized in Table \ref{tab:experiment_results}, 48\% of videos have 100\% agreement (i.e. all 5 independent annotators agreed), 77\% show an 80\% consensus. 
These levels of agreement represent different criteria for labeling a video as (non) creative, and in Section \ref{sec:results} we consider 3 different labelled ground-truth datasets, D-100, D-80, and D-60, based on 100\%, 80\% and 60\% agreement. From Table~\ref{tab:experiment_results} we can also see that 25-30\% of videos were annotated as creative.

\begin{table}[htdp]
\resizebox{\linewidth}{!}{
\begin{tabular}{|c|c||c|c|}
\hline 
Dataset & \% Videos & \# Creative (\%) & \# Non-creative (\%)\\ \hline 
D-60 & 100\% & 1141 (30\%) & 2708 (70\%) \\  
D-80 & 77\% & 789 (27\%) & 2196 (73\%) \\
D-100 & 48\% & 471 (25\%) & 1382 (75\%) \\ \hline
\end{tabular}
}
\caption{Summary of the results of the labeling experiment. D-60: videos with at least 60\% agreement between annotators. D-80: at least 80\% agreement. D-100: 100\% agreement.}
\label{tab:experiment_results}
\end{table}%
\begin{table}[htdp]
\resizebox{\linewidth}{!}{
\begin{tabular}{|c|c|c|c|c|}
\hline 
 & (a) Hashtags & (b) Blogs & (c) Creators & (d) Random \\ \hline 
Creative & 34.05\% & 79.57\% & 27.41\% & 1.88\% \\  \hline  
Non-Creative & 65.95\% & 20.43\% & 72.59\% & 98.12\% \\  \hline  
\end{tabular}}
\caption{Creative vs non-creative videos per sampling strategy, for the D-100 dataset (100\% agreement).}
\label{tab:sources_results}
\end{table}
Table \ref{tab:sources_results} shows the distribution of creative and non-creative videos according to the strategy used to sample the videos. As expected, the videos specifically mentioned in blogs about Vine (b) have the highest proportion of creative videos, while the vast majority of randomly sampled videos (d) are non-creative, justifying the need for our sampling strategies. 
\section{Features for Modeling Creativity} \label{sec:features}
\begin{table*}[t]
\small
\centering
\resizebox{\linewidth}{!}{
	\begin{tabular}{|m{2.5cm}|m{5.2cm}|c|m{10.5cm}|}
\hline
\textbf{Group} &\textbf{Feature} & \textbf{Dim} & \textbf{Description} \\ \hline
\multicolumn{4}{|c|}{\textbf{AESTHETIC VALUE}}\\ \hline
\multicolumn{4}{|c|}{\textit{Sensory Features}}\\ \hline
\textbf{Scene Content} & \textit{Saliency Moments} \cite{sm} &462& Frame content is represented by summarizing the shape of the salient region \\ \hline
& \textit{General Video Properties} &2& \textit{Number of Shots}, \textit{Number of Frames} \\ 
 \textbf{Filmmaking} &\textit{Stop Motion} &1& Number of non-equal adjacent frames \\ 
 \textbf{Technique}&\textit{Loop}  &1& Distance between last and first frame \\ 
&\textit{Movement} &1& Avg. distance between spectral residual \cite{hou2007saliency} saliency maps of adjacent frames\\ 
&\textit{Camera Shake} &1& Avg. amount of camera shake \cite{camerashake} per frame\\ \hline
&\textit{Rule of Thirds} \cite{datta} &3& HSV average value of the inner quadrant of the frame (\textit{H(RoT),S(RoT),V(RoT)})\\ 
\textbf{Composition}  &\textit{Low Depth of Field} \cite{datta} &9& LDOF indicators computed using wavelet coefficients \\ 
\textbf{and Photographic} &\textit{Contrast} \cite{contrast}  &1& Ratio between the sum of max and min luminance values and their difference\\ 
\textbf{Technique}&\textit{Symmetry} \cite{redi2012interestingness}&1& Difference between edge histograms of left and right halves of the image\\ 
&\textit{Uniqueness} \cite{redi2012interestingness}&1& Distance between the frame spectrum and the average image spectrum\\ 
&\textit{Image Order} \cite{order}&2& Order values obtained through Kologomorov \textit{Complexity} and Shannon's Entropy\\ \hline
\multicolumn{4}{|c|}{\textit{Emotional Affect Features}}\\ \hline
 \multirow{4}{*}{\textbf{Visual Affect }} &
\textit{Color Names} \cite{emotions} &9& Amount of color clusters such as red, blue, green, \ldots \\ 
&\textit{Graylevel Contrast Matrix Properties} \cite{emotions} &10& \textit{Entropy, Dissimilarity, Energy, Homogeneity} and \textit{Contrast} of the GLCM matrix \\
&\textit{HSV statistics} \cite{emotions}  &3& Average \textit{Hue, Saturation and Brightness} in the frame\\ \
&\textit{Pleasure, Arousal, Dominance} \cite{valdez1994effects}&3& Affective dimensions computed by mapping HSV values\\ \hline 
\multirow{4}{*}{\textbf{Audio Affect }} 
&\textit{Loudness} \cite{laurier2009exploring} &2&  Overall \textit{Energy} of signal and avg \textit{Short-Time Energy} in a 2-seconds window  \\
&\textit{Mode} \cite{laurier2009exploring} &1& Sums of key strength differences between major keys and their relative minor keys \\ 
&\textit{Roughness} \cite{laurier2009exploring}  &1& Avg of the dissonance values between all  pairs of peak in the sound track spectrum\\ 
&\textit{Rythmical Features} \cite{laurier2009exploring}&2& \textit{Onset Rate} and \textit{Zero-Crossing Rate}\\ \hline
\multicolumn{4}{|c|}{\textbf{NOVELTY}}\\ \hline
  \multirow{2}{*}{\textbf{Novelty} }&
\textit{Audio Novelty} &10& Distance between the audio features and the audio space\\ 
&\textit{Visual Novelty} &40& Distance between the visual features and each visual feature space\\ \hline
 \end{tabular} }
 \caption{Audiovisual features for creativity modeling}
 \label{tab:features}
 \end{table*}
 %
%
In this Section we describe novel and existing features for modeling creative micro-videos, which we group based on the two components of our definition of creative videos: novelty and value. We re-use existing features from computational aesthetics, semantic image analysis, affective image classification, and audio emotions modeling, and propose new features to represent filmmaking technique and novelty. Table \ref{tab:features} summarizes all the features introduced in this section.


%
%
\subsection{Aesthetic Value Features}\label{sec:valuefeatures}
We use a set of features to model the aesthetic value of a video based on two of the three components of aesthetic value identified in Section \ref{sec:creativity}: the \textit{sensory} component and the \textit{emotional} affect of the video. The third, \textit{intellectual}, component is, to the best of our knowledge, not modeled by any existing computational approaches, so we do not model it in this work.
%
%

\subsubsection{Sensory Features}
Sensory features model the raw sensory input perceived by the viewer, which can be approximated by the raw signal output by the video. Such features cover all aspects of the signal, i.e. visual, audio, movement, filmmaking techniques, etc. 
We implement existing features for semantic image classification and aesthetic image analysis, and we design new descriptors to capture the structural characteristics of short-length online videos.\\
\\
\textbf{Video Scene Content.}
We extract the 462-dimensional namely the Saliency Moments feature \cite{sm} from video frames, a holistic representation of the content of an image scene based on the shape of the salient region, which has proven to be extremely effective for semantic image categorization and retrieval.\\ 
\\
\textbf{Composition and Photographic Technique.}
In computational aesthetics, several compositional descriptors describing the photographic and structural properties of images and video frames have been proposed. 
Other features attempt to model the visual theme of images and videos \cite{sparshott1971basic}. 
We use some of the most effective frame-level compositional features, such as the \textit{ Rule of Thirds} and \textit{Low Depth of Field}~\cite{datta}, the \textit{Michelson Contras}t \cite{contrast}, a measure of \textit{Symmetry} \cite{redi2012interestingness}, and a \textit{Uniqueness} \cite{redi2012interestingness} measure indicating the familiarity of the spatial arrangement.  Finally we implement a feature describing the \textit{Image Order} using information theory-based measurements~\cite{order}.\\
\\
\textbf{Filmmaking Technique Features.}
We design a set of new features for video motion analysis, inspired by movie theory and tailored to model the videomaking techniques of short on-line videos.
\\ \textit{General Video Properties.} We compute the number of frames $N_f$ and the number of shots $N_s$ in the video. In the current setting, the number of frames is a proxy for frame rate, as almost all videos are exactly 6 seconds in length, whereas the frame rate tends to vary. 
\\\textit{Stop Motion.}   
Many popular creative short videos are stop-motion creations, where individual  photos are concatenated to create the illusion of motion. In such videos 
the frequency of changes in the scene is lower than traditional videos.  We capture this technique by computing the Euclidean distance $\delta (F_i, F_{i+1})$ between the pixels of neighboring frames $F_i$ and $F_{i+1}$ and then retaining as a stop motion measure $S$ the ratio between $N_f$ and the number of times such difference is not null (the scene is changing), namely 
    \begin{equation}
S=\frac{N_f}{1+\sum_{i=1}^{N_f-1}sgn(\delta(F_i,F_i+1))}.
  \end{equation} 
\\\textit{Loop.} Many popular videos in Vine are shared with the hashtag \#loop. A looping video carries a repeatable structure that can be watched repeatedly without perceiving where the beginning/end of the sequence is. To capture this, we  compute the distance between the first and the last frames of the video, namely  $ L=\delta(F_1,F_{N_f}) $
\\\textit{Movement.} similar to previous works, \cite{Chung:EECS-2012-172, bhattacharya2013towards}, we compute the amount of motion in a video using a feature that can describe the speed of the main objects in the image regardless of their size. We first compute a saliency map of each frame 
 and then retain, as a movement feature, the average of the distances between the maps of neighboring frames:
     \begin{equation}
M=1/N_f\sum_{i=1}^{N_f-1}\delta(SM(F_i),SM(F_{i+1})) 
  \end{equation}
where $SM(\cdot)$ is the saliency map computed on the frame using the Spectral Residual technique \cite{hou2007saliency}.
\\\textit{Camera-Shake.} Typical micro-videos are not professional movies,  and often contain camera shake introduced by handheld mobile phone cameras. Artistic video creators, however, often carefully produce their videos, avoiding camera-shake. We compute the average amount of camera shake in each frame using an approach based on the directionality of the Hough transform computed on image blocks \cite{camerashake}. 
 %
\begin{figure*}[t]
  \centering
    \includegraphics[width=\linewidth]{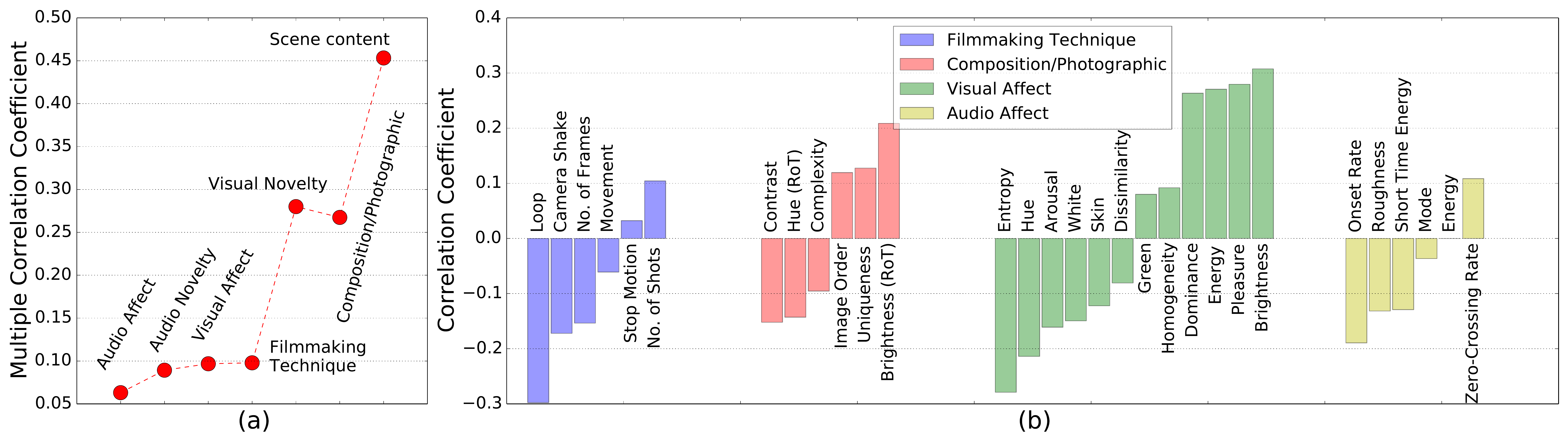}
  \caption{Analysis of the most relevant features and components for video creativity prediction}
  \label{analysis}
\end{figure*}
%
\subsubsection{Emotional Affect Features}
In this section we separately introduce sets of visual and audio features known to correlate with emotional affect.
\\
\\
\textbf{Visual Affect.}
We extract a set of frame level affective features, as implemented by Machajdik \& Hanbury \cite{emotions}, namely \textit{Color names},  \textit{Graylevel Contrast Matrix (GLCM) properties},  \textit{Hue, Saturation and Brightness} statistics, \textit{Level of Detail}, and the \textit{Pleasure}, 
\textit{Arousal}, 
and \textit{Dominance} 
values computed from HSV values \cite{valdez1994effects} . 
\\
\\
\textbf{Audio Affect.}
Inspired by Laurier et al \cite{laurier2009exploring}, we implement, using the MIRToolbox \cite{lartillot2007matlab}, a number features for describing audio emotions, collecting them a 6-dimensional feature vector. We describe the sound \textit{Loudness}, the overall volume of the sound track, its  \textit{Mode} (indicating if the sound in the Major or Minor mode), the audio \textit{Roughness} (dissonance in the sound track), and \textit{{Rythmical Features}} describing abrupt rhythmical changes in the audio signal.

\subsection{Novelty}
The novelty of an artifact can be represented by its distance from a set of other artifacts of the same type. One way to compute such distance is to first divide the attribute space into $K$ clusters, and then calculate the distance the between the artifact and its nearest cluster~\cite{maher2010evaluating}. 
In our approach, we compute an improved novelty feature that takes into account the distances between the artifact attribute and \textit{all the clusters} in the attribute space, thus measuring not only the distance to the most similar element, but the detailed position of the attribute in the space.

We measure novelty for both the visual and the audio channel of the video, using as attributes the aesthetic values features from Section \ref{sec:valuefeatures}. 
We take a random set of  videos, independent of our annotated corpus, and extract the 4 groups of visual attributes (\textit{Scene Content (SC)}, \textit{Filmmaking Techniques}, \textit{Composition and Photographic Technique} and \textit{Visual Affect}), and the \textit{Audio Affect} attributes.
We cluster the space resulting from each attribute into $10$ clusters using K-means, obtaining 40 clusters for the visual attributes (10 clusters each for 4 attributes) and 10 for the audio attribute. 

To calculate the novelty score for a given video, we extract the visual and audio attributes, and we then compute the \textbf{Audio  Novelty} as the collection of the distances between the \textit{Audio Affect} attribute of the video and all the clusters of the corresponding space (giving a 10 dimensional feature). Similarly, we compute the video \textbf{Visual  Novelty} as the set of distances between each visual attribute of the video and the corresponding cluster set (40 dimensions).

\section{Experimental Results}\label{sec:results}
In this Section we explore the extent to which audio-visual features correlate with creative video content, and then evaluate the approaches for creative video classification.

\subsection{What Makes a Video Creative?}
To analyze which features correlate most with creative micro-videos, we consider videos with 100\% agreement (i.e. D-100 from Table \ref{tab:experiment_results}), as we are interested in the correlations for the cleanest version of our dataset. We extract 7 groups of features for each video: \textit{Scene Content, Composition/Photographic Technique, Filmmaking Technique, Visual Emotional Affect, Audio Emotional Affect, Visual Novelty,} and \textit{Audio Novelty.} For frame-level features, we consider the features of middle frame of the video. 

We first analyze to what extent each group of features correlates with video creativity, using the \textit{Multiple Correlation Coefficient ($MPC$)}, which measures how well a multidimensional variable fits a monodimensional target variable, given the reconstructed signal after regression.
In our context, the elements of the multidimensional variable are the individual features within a feature group.

In Figure \ref{analysis}(a), we report MPC values for all features, testing how well each of our groups of features fits the label vector of our data. 
The results show that \textit{Scene Content} is most strongly correlated with creative videos, followed by \textit{Composition/Photographic} features and \textit{Video Novelty}, showing that both novelty and aesthetic value features are crucial for understanding creativity. Emotional affect features are less strongly correlated than sensory features, suggesting that the raw sensory output of the audiovisual stream is more useful than features that attempt to model emotional affect. The correlation for audio features is much lower than for visual features: it seems likely that micro-video authors and annotators both place more emphasis on the visual aspect.

To determine, within each group of features, the most important individual features, we calculate the Pearson correlation coefficient $\rho$  between individual features and creative labels. 
For this analysis, we exclude the \textit{Scene Content} features, whose elements are non-separable, and the \textit{Novelty} features, which require all features to represent the position of the video in the attribute space. 

In  Figure \ref{analysis}(b) we show the $\rho$ values for the most highly correlated features. 
Among the \textit{Composition/Photographic Technique} and \textit{Visual Affect} features, we can see that creative videos strongly correlate with visual uniformity (positive correlation with  \textit{Energy}), and order (negative correlation with \textit{Entropy} and the \textit{Complexity} measure), suggesting that videos with homogenous frames are more likely to be perceived as creative. 
	Moreover, a negative correlation with the \textit{Hue} statistic in both the whole image (\textit{Visual Affect - Hue}) and in the inner quadrant (\textit{Composition/Photographic - Hue (RoT)}) and a highly positive $\rho$ for \textit{Visual Affect - Brightness} and \textit{Composition/Photographic - Brightness (RoT)} shows that creative videos are related to warm and bright colors. The \textit{Uniqueness} feature shows a positive $\rho$, indicating videos with a less familiar layout are more likely to be labeled as creative. Surprisingly, important properties for visual aesthetics such as \textit{Symmetry} ($\rho=0.04$) and \textit{Low Depth of Field}   ($\rho=0.02$)  do not play a key role for creative video detection (these are not shown in Figure \ref{analysis}(b), which only includes features with a high correlation). Also, \textit{Skin} color (Visual Affect) is negatively correlated with creativity, showing that the presence of people is not so common in creative videos, unlike most other popular videos.
	
Creative videos are associated with highly pleasant emotions,  and dominant, non-overwhelming, controllable emotions ($\rho=0.24$ for \textit{Pleasure} and $\rho=0.23$ for \textit{Dominance}). 
In terms of \textit{Audio Affect}, \textit{Onset Rate} and loudness (\textit{Short Time Energy}) are negatively correlated with creativity, meaning that less frenetic, low-volume sound is preferred.

Regarding \textit{Filmmaking Techniques}, the most highly correlated feature is the presence of the loop technique, a common form of expression in micro-videos (it is negatively correlated because a high score indicates low likelihood of a loop). \textit{Camera Shake}, which can be seen as an inverse quality measure, is a negative indicator of creativity, suggesting that more `polished' videos are likely to be seen as creative.

As we can see from our findings, creativity involves a variety of dimensions beyond photographic beauty. Photographic features traditionally used in visual aesthetic frameworks \cite{datta,interesting} are equally or even less correlated with creativity than other, complementary, features. Moreover, we can see that single \textit{Filmmaking Technique} features such as \textit{Loop}, or \textit{Audio Affect} features such as \textit{Onset Rate}, are better indicators of creativity, compared to single photographic features.

\subsection{Classifying Creative Videos}
We now evaluate methods to automatic classify videos as creative, using the same features.
\\
\\
\textbf{Experimental Setup.} In Table~\ref{tab:experiment_results} in Section \ref{sec:dataset} we described three different versions of our annotated corpus, based on 60\%, 80\% and 100\% per-video annotation agreement. These datasets give a natural tradeoff between dataset size and label accuracy. 
To measure the effect of this tradeoff on classification, we report results for each dataset.

For each version of the corpus we use 2/3 of the positive examples for training, and the rest for testing. For training and testing, we subsample an equal amount of negative examples, to ensure a balanced set. We train a separate Support Vector Machine with Radial Basis Function (RBF) kernel for each of the 7 groups of features. For groups of features that are calculated for a single video frame, at the training stage we sample 12 frames for the video, and create a separate training instance for each sampled frame, each given the label of the parent video.
We use the trained models to classify the creative videos in the test set. For each video, the classifier outputs a label and a classification score. 
For the frame-level features, we sample 12 frames as in training, classify each, and retain as overall classification of the video the rounded average of the single frame scores. We use classification accuracy as our evaluation measure.

For the novelty features, we use 1000 non-annotated videos for the clustering. To check that this number does not introduce any bias in our experiment, we re-compute clustering on an increasing number of videos, from 500 to 5000, and obtained similar results as those presented in Table \ref{tab:results}. 

To test the complementarity of the groups of features and the improvement obtained by combining them, we also combine the classification scores of different classifiers using the  \textit{median} value of the scores of all the classifiers, previously shown to perform well for score aggregation \cite{kittler1998combining}.
\\
\\
\textbf{Results.} The classification results are shown in Table \ref{tab:results}. Similar to the correlations, we can see that the best feature group is \textit{Composition/Photographic Technique}, with 77\% accuracy (D-100 dataset), followed by \textit{Scene Content} and \textit{Filmmaking Technique} features. We can also see that Emotional Affect features are outperformed by Sensory features. Our new, 6-dimensional, \textit{Video Technique} feature achieves comparable classification accuracy to the performance of the 462 dimension \textit{Scene content} feature. Combining emotional and sensory features improves classification accuracy to  79\%, showing the complementarity of these features.


\begin{table}[htbp]
\resizebox{1\linewidth}{!}{
\begin{tabular}{|p{6cm}|c|c|c|}
\thickhline
  \multirow{2}{*}{\textbf{Feature}}
  &  \multicolumn{3}{c|}{\textbf{Accuracy}}\\\cline{2-4}
 &  \textbf{D-60} &  \textbf{D-80} & \textbf{D-100} \\ \thickhline
  \textbf{Aesthetic Value} &  &  &  \\ \thickhline
 \textit{Sensory Features} &  &  &  \\ 
 Scene Content & 0.67 & 0.69 & 0.74 \\
 Filmmaking Techniques & 0.65 & 0.69 & 0.73 \\
Composition \& Photographic Technique & 0.67 & 0.74 & \textbf{0.77}\\
All Sensory Features & 0.69 & \textbf{0.75} & {0.77} \\ 
\textit{Emotional Affect Features} &  &  &  \\ 
Audio Affect & 0.59 & 0.53 & 0.67 \\
Visual Affect & 0.65 & 0.66 & 0.66 \\
All Emotional Affect Features  & 0.62 & 0.56& \textbf{0.71} \\
\textbf{All Aesthetic Value Features}  & 0.68 & 0.72 & \textbf{0.79} \\\thickhline
\textbf{Novelty} &  &  & \\ \thickhline
Audio & 0.58 & 0.58 & 0.63 \\
Visual & 0.63 & 0.67 & \textbf{0.74} \\
Audio + Visual Novelty & 0.59 & 0.63 & 0.69 \\\thickhline
\textbf{Novelty + Aesthetic Value} & 0.69 & 0.73 & \textbf{0.80} \\\thickhline
 \end{tabular} }
 \caption{Prediction results for value and novelty features} \centering
 \label{tab:results}
 \end{table}

Although the \textit{Novelty} features carry some discriminative power for creative video classification, Aesthetic Value features are still more discriminative. However, when we combine novelty and value features, we can 
see their complementarity, with the classification accuracy increased from 79\% to 80\% for the D-100 dataset. 

Overall, we can notice the importance of using a diversity of features for creativity prediction, since classifiers based on traditional photographic features  or generic scene features, typical of visual aesthetic frameworks, benefit from the combination with other cues, justifying a tailored framework for creative video classification. 

Finally, we can also see that the quality of the annotations is crucial: classification accuracy is always much higher for the cleanest dataset, D-100, even though this dataset is only 60\% the size of the D-80 dataset, and less than half the size of the D-60 dataset.


\section{Conclusions} \label{sec:conclusions}
In this paper, we study creativity in short videos, or \textit{micro-videos}, shared in online social media platforms such as Vine or Instagram. Defining creative videos as videos that are \textit{novel} (\ie, surprising, unexpected) and have \textit{aesthetic value}, we run a crowdsourcing experiment to label more than 3,800 \textit{micro-videos} as creative or non-creative. We obtain a high level of inter-annotator agreement, showing that, with appropriate guidelines, it is possible to collect reliable annotations for a subjective task such as this. From this annotation we see that a small, but not insignificant, 1.9\% of randomly sampled videos are labeled as creative.

We propose a number of new and existing computational features, based on \textit{aesthetic value} and \textit{novelty}, for modeling creative \textit{micro-videos}. We show that groups of features based on scene content, video novelty, and composition and photographic technique are most correlated with creative content. We show that specific features measuring order or uniformity correlate with creative videos, and that creative videos tend to have warmer, brighter colors, and less frenetic, low volume sounds. Also, they tend to be associated with pleasant emotions, and dominant, non-overwhelming, controllable emotions. Loop and Camera Shake features, specifically designed for modeling creativity in \textit{micro-videos}, also show high correlation with creativity. Several features traditionally associated with beauty or interestingness show low correlations with creative \textit{micro-video}, underlining the difference between creativity and those concepts. Specifically, skin color, symmetry and low depth, which  are widely used in modeling beauty and interestingness, are not correlated with creative \textit{micro-videos}.

Finally, we evaluate approaches to the automatic classification of creative \textit{micro-videos}. We show promising results overall, with a highest accuracy of 80\% on a balanced dataset. The best results are achieved when we combine novelty features with aesthetic value features, showing the usefulness of this twofold definition of creativity. We also show that high quality ground truth labels are essential to train reliable models of  creative micro-videos.

In future work, we plan to enlarge the set of features for modeling creativity. We will design features to model the intellectual aspect of aesthetic value through semantic visual cues such as specific visual concept detectors. Moreover, we plan to include non-audiovisual cues such as the metadata related to the video (tags, tweets, user profile), the comments about it, and its' popularity in the social media community. 

Furthermore, we would like to apply our model, or a modified version of it, to other micro-video platforms and also to a broader spectrum of multimedia content, such as images and longer videos, \etc, and to study the differences and commonalities between their creative features. 


{\small
\bibliographystyle{ieee}
\bibliography{cvpr2014}

\begin{thebibliography}{10}\itemsep=-1pt

\bibitem{camerashake}
http://www.cs.bgu.ac.il/~ben-shahar/teaching/computational-vision/studentprojects/icbv121/icbv-2012-1-kerendamari-bensimandoyev/index.php.

\bibitem{bhattacharya2013towards}
S.~Bhattacharya, B.~Nojavanasghari, T.~Chen, D.~Liu, S.-F. Chang, and M.~Shah.
\newblock Towards a comprehensive computational model foraesthetic assessment
  of videos.
\newblock In {\em Proceedings of the 21st ACM international conference on
  Multimedia}, pages 361--364. ACM, 2013.

\bibitem{bhattacharya2010framework}
S.~Bhattacharya, R.~Sukthankar, and M.~Shah.
\newblock A framework for photo-quality assessment and enhancement based on
  visual aesthetics.
\newblock In {\em ACM Multimedia}, pages 271--280, 2010.

\bibitem{Chung:EECS-2012-172}
S.~Chung, J.~Sammartino, J.~Bai, and B.~A. Barsky.
\newblock Can motion features inform video aesthetic preferences?
\newblock Technical Report UCB/EECS-2012-172, EECS Department, University of
  California, Berkeley, Jun 2012.

\bibitem{datta}
R.~Datta, D.~Joshi, J.~Li, and J.~Wang.
\newblock Studying aesthetics in photographic images using a computational
  approach.
\newblock In {\em IEEE ECCV}, pages 288--301, 2006.

\bibitem{contrast}
M.~Desnoyer and D.~Wettergreen.
\newblock Aesthetic image classification for autonomous agents.
\newblock In {\em ICPR}, 2010.

\bibitem{interesting}
S.~Dhar, V.~Ordonez, and T.~Berg.
\newblock High level describable attributes for predicting aesthetics and
  interestingness.
\newblock In {\em IEEE CVPR}, pages 1657--1664.

\bibitem{higgins1999applying}
L.~F. Higgins.
\newblock Applying principles of creativity management to marketing research
  efforts in high-technology markets.
\newblock {\em Industrial Marketing Management}, 28(3):305--317, 1999.

\bibitem{hou2007saliency}
X.~Hou and L.~Zhang.
\newblock Saliency detection: A spectral residual approach.
\newblock In {\em IEEE CVPR}, pages 1--8, 2007.

\bibitem{isola2011makes}
P.~Isola, J.~Xiao, A.~Torralba, and A.~Oliva.
\newblock What makes an image memorable?
\newblock In {\em IEEE CVPR}, pages 145--152. ACM, 2011.

\bibitem{kant1987critique}
I.~Kant and W.~S. Pluhar.
\newblock {\em Critique of judgment}.
\newblock Hackett Publishing, 1987.

\bibitem{ke2006design}
Y.~Ke, X.~Tang, and F.~Jing.
\newblock The design of high-level features for photo quality assessment.
\newblock In {\em IEEE CVPR}, pages 419--426, 2006.

\bibitem{kittler1998combining}
J.~Kittler, M.~Hatef, R.~P. Duin, and J.~Matas.
\newblock On combining classifiers.
\newblock {\em IEEE PAMI}, 20(3):226--239, 1998.

\bibitem{lartillot2007matlab}
O.~Lartillot and P.~Toiviainen.
\newblock A matlab toolbox for musical feature extraction from audio.
\newblock In {\em International Conference on Digital Audio Effects}, pages
  237--244, 2007.

\bibitem{laurier2009exploring}
C.~Laurier, O.~Lartillot, T.~Eerola, and P.~Toiviainen.
\newblock Exploring relationships between audio features and emotion in music.
\newblock {\em ESCOM}, 2009.

\bibitem{luo2011content}
W.~Luo, X.~Wang, and X.~Tang.
\newblock Content-based photo quality assessment.
\newblock In {\em ICCV}, pages 2206--2213. IEEE, 2011.

\bibitem{emotions}
J.~Machajdik and A.~Hanbury.
\newblock Affective image classification using features inspired by psychology
  and art theory.
\newblock In {\em Multimedia}, pages 83--92. ACM, 2010.

\bibitem{maher2010evaluating}
M.~L. Maher.
\newblock Evaluating creativity in humans, computers, and collectively
  intelligent systems.
\newblock In {\em Proceedings of the 1st DESIRE Network Conference on
  Creativity and Innovation in Design}, pages 22--28. Desire Network, 2010.

\bibitem{marchesotti2011assessing}
L.~Marchesotti, F.~Perronnin, D.~Larlus, and G.~Csurka.
\newblock Assessing the aesthetic quality of photographs using generic image
  descriptors.
\newblock In {\em IEEE ICCV}, pages 1784--1791, 2011.

\bibitem{mason2011}
W.~Mason and S.~Suri.
\newblock Conducting behavioral research on amazon's mechanical turk.
\newblock {\em Behavior Research Methods}, 44(1):1--23, June 2011.

\bibitem{moorthy2010towards}
A.~K. Moorthy, P.~Obrador, and N.~Oliver.
\newblock Towards computational models of the visual aesthetic appeal of
  consumer videos.
\newblock In {\em IEEE ECCV}, pages 1--14. 2010.

\bibitem{mumford2003have}
M.~D. Mumford.
\newblock Where have we been, where are we going? taking stock in creativity
  research.
\newblock {\em Creativity Research Journal}, 15(2-3):107--120, 2003.

\bibitem{murray2012ava}
N.~Murray, L.~Marchesotti, and F.~Perronnin.
\newblock Ava: A large-scale database for aesthetic visual analysis.
\newblock In {\em IEEE CVPR}, pages 2408--2415, 2012.

\bibitem{newell1959processes}
A.~Newell, J.~Shaw, and H.~A. Simon.
\newblock {\em The processes of creative thinking}.
\newblock Rand Corporation, 1959.

\bibitem{nishiyama2011aesthetic}
M.~Nishiyama, T.~Okabe, I.~Sato, and Y.~Sato.
\newblock Aesthetic quality classification of photographs based on color
  harmony.
\newblock In {\em IEEE CVPR}, pages 33--40, 2011.

\bibitem{sm}
M.~Redi and B.~Merialdo.
\newblock Saliency moments for image categorization.
\newblock In {\em ACM ICMR}, 2011.

\bibitem{redi2012interestingness}
M.~Redi and B.~Merialdo.
\newblock Where is the interestingness? retrieving appealing videoscenes by
  learning flickr-based graded judgments.
\newblock In {\em ACM Multimedia}, pages 1363--1364, 2012.

\bibitem{order}
J.~Rigau, M.~Feixas, and M.~Sbert.
\newblock Conceptualizing birkhoff's aesthetic measure using shannon entropy
  and kolmogorov complexity.
\newblock {\em Computational Aesthetics in Graphics, Visualization, and
  Imaging}, 2007.

\bibitem{sparshott1971basic}
F.~Sparshott.
\newblock Basic film aesthetics.
\newblock {\em Journal of Aesthetic Education}, 5(2):11--34, 1971.

\bibitem{valdez1994effects}
P.~Valdez and A.~Mehrabian.
\newblock Effects of color on emotions.
\newblock {\em Journal of Experimental Psychology}, 123(4):394, 1994.

\bibitem{weisberg1993creativity}
R.~W. Weisberg.
\newblock {\em Creativity: Beyond the myth of genius}.
\newblock 1993.

\bibitem{yao2012oscar}
L.~Yao, P.~Suryanarayan, M.~Qiao, J.~Z. Wang, and J.~Li.
\newblock Oscar: On-site composition and aesthetics feedback through exemplars
  for photographers.
\newblock {\em IJCV}, 96(3):353--383, 2012.

\end{thebibliography}
}

\end{document}